\documentclass[sigconf]{acmart}

\AtBeginDocument{%
  }

\copyrightyear{2025}
\acmYear{2025}
\setcopyright{cc}
\setcctype{by}
\acmConference[CHI '25]{CHI Conference on Human Factors in Computing Systems}{April 26-May 1, 2025}{Yokohama, Japan}
\acmBooktitle{CHI Conference on Human Factors in Computing Systems (CHI '25), April 26-May 1, 2025, Yokohama, Japan}\acmDOI{10.1145/3706598.3713854}
\acmISBN{979-8-4007-1394-1/25/04}

\usepackage{hyperref}
\usepackage{footnote} 
\usepackage{enumitem}
\usepackage{multirow}
\usepackage{tabularx}

\begin{document}

\title{Copying style, Extracting value: Illustrators’ Perception of AI Style Transfer and its Impact on Creative Labor}

\author{Julien Porquet}
\affiliation{
  \institution{University of Cambridge}
  \city{Cambridge}
  \country{United Kingdom}
}
\email{jp980@cam.ac.uk}

\author{Sitong Wang}
\affiliation{
  \institution{Columbia University}
  \city{New York}
  \state{NY}
  \country{USA}
}
\email{sw3504@columbia.edu}

\author{Lydia B. Chilton}
\affiliation{
  \institution{Columbia University}
  \city{New York}
  \state{NY}
  \country{USA}
}
\email{chilton@cs.columbia.edu}

\renewcommand{\shortauthors}{Porquet et al.}

\begin{abstract}
Generative text-to-image models are disrupting the lives of creative professionals. 
Specifically, illustrators are threatened by models that claim to extract and reproduce their style. 
Yet, research on style transfer has rarely focused on their perspectives. 
We provided four illustrators with a model fine-tuned to their style and conducted semi-structured interviews about the model’s successes, limitations, and potential uses. 
Evaluating their output, artists reported that style transfer successfully copies aesthetic fragments but is limited by content-style disentanglement and lacks the crucial emergent quality of their style. 
They also deemed the others’ copies more successful. 
Understanding the results of style transfer as “boundary objects,” we analyze how they can simultaneously be considered unsuccessful by artists and poised to replace their work by others.
We connect our findings to critical HCI frameworks, demonstrating that style transfer, rather than merely a Creativity Support Tool, should also be understood as a supply chain optimization one.

\end{abstract}

\ccsdesc[500]{Human-centered computing~Empirical studies in HCI}

\keywords{Generative AI, Style Transfer, Creative Labor, Boundary Objects, Capitalism}

\maketitle

\section{Introduction}

\begin{figure*}
\centering
\includegraphics[width=0.9\textwidth]{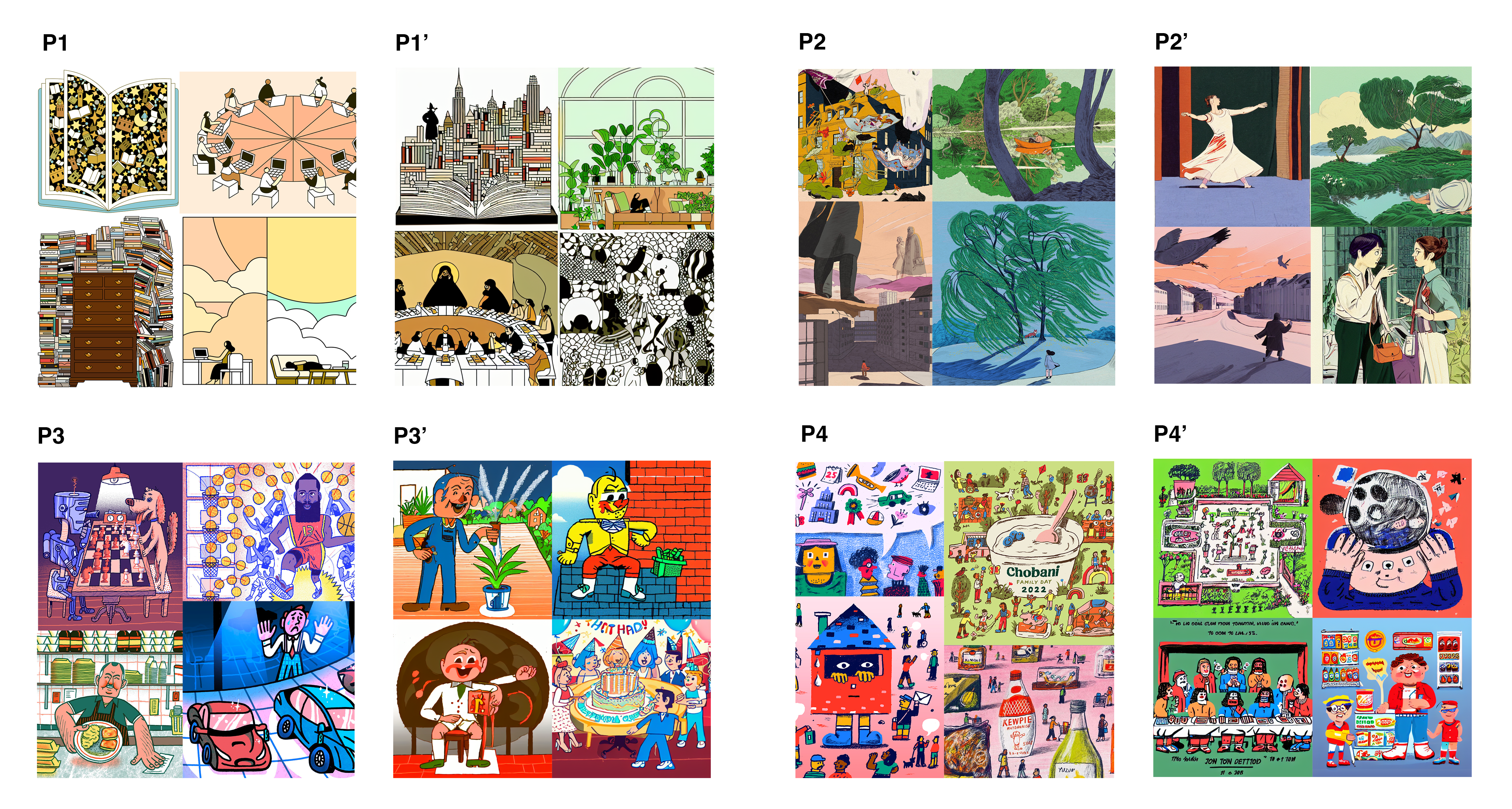}
\caption{Side by side of participant's work (Pn) and style transfer results (Pn')}
\label{fig:overview1}
\Description{4 pairs of 8 illustrations are presented on a grid. Each pair comprises four illustrations created by one of the participants on the left-hand side and four illustrations generated by our model on the right-hand side. Above the illustrator’s four images is their participant code (Pn), and above the generated images is Pn’. Each pair gives an overview of general stylistic similarity between the illustrators’ work and their models.}
\end{figure*}

For commercial artists like illustrators working in editorial, advertising, and branding contexts, cultivating a singular style is crucial to their creative identity and their ability to make a living. 
Participating in their artistic and economic success, style is a rich and complex feature of their work, and by extension of the creative economy to which they are essential contributors. 
Meanwhile, generative AI models, often trained on these illustrators’ work, have enabled the generation of new images in many styles.
In particular, style transfer---the technique of extracting an artistic style from an image and applying it to another---has long fascinated computer science researchers. 
The emerging use of style transfer to replicate illustrators’ styles, often without their consent, raises significant ethical concerns \cite{AIart2,scienceAIart}. 

Yet, despite the central role of style in their work, computer scientists have not fully defined the concept or its relation to specific art history approaches or industry realities, highlighting the need for critical examination (we note the seminal exception of \cite{elgammal2018shape}). 
This lack of specificity is characteristic of the “view from nowhere” approach to technology criticized by feminist scholars \cite{haraway}, and is potentially harmful to artists.
Equating what is generated by style transfer to the rich and complex reality of artists’ style has contributed to consequential narratives of artists’ obsolescence and furthered their alienation from artistic production.
Instead, style is best understood specifically, located in the political economy of illustration as a profession. Foregrounding illustrators’ perspectives, this paper asks two questions:
\begin{itemize}
    \item RQ1: How do illustrators perceive style transfer results?
    \item RQ2: Who do these results benefit within the political economy of creative labor?
\end{itemize}

\begin{figure*}
\centering
\includegraphics[width=0.9\textwidth]{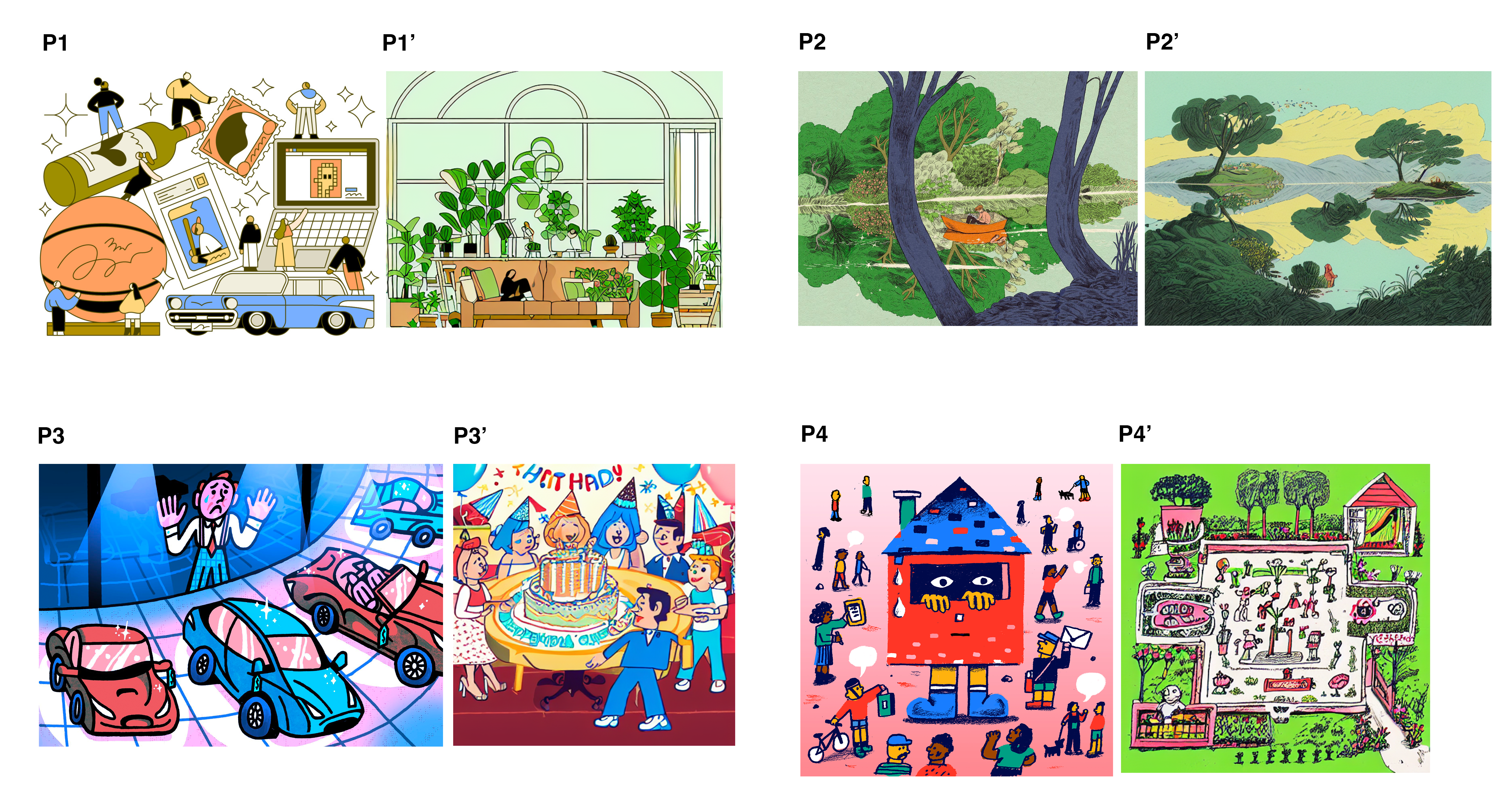}
\caption{Detailed side-by-side of participant's work (Pn) and style transfer results (Pn')}
\label{fig:overview2}
\Description{ pairs of 2 images are presented on a grid. Each pair is composed of 2 illustrations. The first image of each pair is created by one of the participants on the left-hand side, and four illustrations are generated by our model on the right-hand side. Above the illustrator’s four images is their participant code (Pn), and above the generated images is Pn’. Each pair gives an overview of general stylistic similarity between the illustrators’ work and their models.
}
\end{figure*}

We invited four illustrators to generate images with a diffusion model fine-tuned to their own work to investigate these questions (see Figures \ref{fig:overview1} and \ref{fig:overview2}). 
We then conducted semi-structured interviews about their evaluation of the quality, similarity, and usefulness of the output compared to their practice. 
They were also asked to evaluate the success of the other participants’ models. 
These results were contextualized in a larger-scale ethnographic fieldwork conducted by one of the authors with the same illustrators.

To understand how researchers, artists, and their clients can use the same concept while referring to various experiences, we mobilize the concept of “boundary objects”, introduced by Star and Griesemer \cite{star1989institutional}.
Boundary objects share common features across various communities yet also adapt to local contexts, making them useful for understanding human-computer interactions \cite{boundaryobject1, boundaryobject2, boundaryobject3}.
Conceptualizing style as a boundary object enables us to anchor the performance of style transfer within particular contexts while understanding how its presence across different social worlds can either facilitate or hinder collaboration.

Our findings suggest that the version of style generated by style transfer differs from what illustrators consider as style.
We found that style transfer successfully reproduces isolated elements of the artists' style, such as colors, shading, or textures. Still, the result never exceeded the sum of these parts to be evaluated as entirely successful. 
The content-style disentanglement on which style transfer relies was also at odds with how illustrators worked with style. They reported that to convey style, one has to do so through both semantic and aesthetic elements.
Reflecting on their process, participants have shared how style, rather than a set plan to follow and apply to future works, was negotiated during the creative process, guided by personal taste, and characterized by difference and surprise rather than similarity and reproduction.
Finally, we found that artists were more likely to positively evaluate other participants’ results than their own, pointing out how style evaluation is not uniform across evaluators but socially located.

We contextualize these findings within the political economy of illustration work to show how the design and limitations of style transfer create boundary objects that hinder collaboration with artists but facilitate the extraction of value from their work by their clients.
By reproducing a decontextualized, shallow, and predictable copy of style, we argue that rather than a creativity support tool (CST), style transfer is best understood as a supply-chain optimization tool.

\section{Background}

\subsection{Style in illustration}
Borrowing from art historian Meyer Schapiro \cite{Schapiro1994-SCHTAP-18}, we can define style, provisionally, as “the constant form---and sometimes the constant elements, qualities, and expression---in the art of an individual or a group.” 
In art history, style is a category used to classify artworks by one artist or group, often retrospectively \cite{arnheim, panofsky1995three}. 
For contemporary illustrators, on the other hand, style is a “generative principle” \cite{wilf2013media} negotiated at the moment of the creative act and which guides future decisions.
At the intersection of identity and aesthetics, art and economy, making and seeing, style is central to illustrators’ creative practice and ability to make a living.

\subsubsection{Artistic value of style}
Illustrators spend years cultivating a “signature style.” 
In a blog post~\cite{marloes}, picture book illustrator Marloes De Vries articulates the relationship of style to identity as such: 
\begin{quote}
\textit{
“[...] it’s already in you and the more you practice it, the more it forms. The illustration style that will surface is a combination of your natural way of drawing[...], what you are interested in (the subjects you draw) and the technique you choose to work in. Those combined are your style.”
}
\end{quote}

This notion of style as “already in you” is anchored in romantic ideologies of creativity as individual self-expression \cite{rose1993authors,woodmansee2017genius} and plays a central role in establishing the value of artworks as signs of an authentic artist’s inner self \cite{wilf2014semiotic}. 

While in part unique and personal to them, styles are also profoundly social objects. 
Artists’ styles develop through engaging with various historical and contemporary influences, and individual styles can be grouped into larger categories identifying subcultures or temporary trends. 
For the illustrators we interviewed, their style was about social connection with the artists who came before and contemporary ones. 
Yet, the social life of style also involves a degree of anxiety around plagiarism for artists, an issue intimately tied to generative AI that some call a “plagiarism machine” \cite{plagiarismmachine,plagiarismmachine2}.

\subsubsection{Economic value of style}
In addition to being central to artistic identity, style is a valuable commodity in late capitalist economies, where consumers craft their identity by purchasing branded (stylized) goods \cite{pang2009labor,chumley2016creativity}.
Freelance illustrators are hired by magazines, advertising or branding studios for their style to contribute articles, campaigns or brands. 
As a unique sign of their makers, style is also involved in competitiveness, differentiating illustrators in a saturated market. 
Style thus has tremendous economic value, and regulating, reproducing, and capitalizing on style is a central concern for the creative industry.

Despite this importance, styles are also not protected by intellectual property laws \cite{fitzpatrick1992hazards,brownlee1993safeguarding}, which makes them highly volatile commodities \cite{pang2009labor}. 
As most of the major companies involved in creating generative models are currently facing lawsuits for copyright infringement led by several illustrators \cite{AIlawsuits}, style proves to be a crucial, albeit unruly, object to examine in the interaction between humans and machines. 

At the intersection of artistic, economic, and legal ideologies, the concept of style is central to many stakeholders in the creative industry for different reasons. 
We now zoom in on illustrators' main challenges in retaining control over their work (and style) in the creative industry, challenges that constitute the context in which style transfer unfolds. 

\subsection{The political economy of style}
\label{sec:political_economy_style}
Illustration has been a central component of companies' branding and advertising projects, with many illustrators making most of their income from these sources (vs. editorial work, which is way less paid). 
Branding and advertising studios act as strategists and producers for these projects and as brokers of contractual artistic work (hiring illustrators, photographers, etc.). 
Studios take a client’s brief, create a concept, and source various artists whose work will help visually convey the client's tone, ideas, and voice. 

While designers usually aim to work “from the ground up” and create visuals motivated by systems and concepts, they often have to deal with clients’ superficial understanding of design and illustration, tight budgets, and short turnarounds, leading them to focus on aesthetics first \cite{Brandidentity}. 
These pressures motivate several strategies for objectifying, extracting, and appropriating artists' styles. 
We outline three of these strategies here. 

\textbf{(1) Design strategy}: Moodboards are collections of images pulled from the internet, often without a trace of who made them, grouped to convey a stylistic direction for a client. 
Design critic Elizabeth Goodspeed \cite{moodboard} showed how moodboards lead to derivative work, as “these visual cues operate on a surface level, rendering form without function.” 
This divorce of form from function echoes our findings about the content-style disentanglement (section \ref{sec:content-style-disentanglement}), showing how such logic fits within a particular political economy of creative work. In addition, moodboards erase the identity of artists by extracting their style from its context without crediting them. 
This results in conceptualizing styles as natural resources ready for the taking rather than the product of human labor.

\textbf{(2) Labor strategy}: Once styles are extracted from their creators, they can be applied superficially elsewhere by other actors. 
This has happened routinely, long before generative models, in studios seeking to save money by producing illustrations “in-house.” 
In doing so, employees (whose work is not protected by copyright) create work in any style at a fraction of the cost of a freelance illustrator. 
In this process, styles become fragmented, isolated elements, with designers taking discreet elements from various images, much like the fragmented results our participants described in section \ref{sec:sucessful-fragments}.
This practice is, essentially, an analog precursor of style transfer. 

\textbf{(3) Legal strategy}: Illustrators are freelance creative workers whose work is protected by copyright law (note that their style is not).
Following this, the standard legal contract between an illustrator and a client is a licensing agreement through which the artist licenses their work for a given use, period, and territory. 
Yet, freelancers see the rise of “work-for-hire” contracts, which means that an artist never owns the copyright of their work and cannot display the work in their portfolio \cite{workforhire}. 
These legal strategies extract value from artists' work while furthering their alienation.
Understanding the legal context in which a technology like style transfer operates is crucial, as several AI companies are currently being sued for copyright infringement \cite{AIlawsuits}.

These strategies have (1) lowered the quality and originality of artistic work, (2) undermined the role of individual creators, and (3) prevented artists from retaining authorship in their work.
They outline the challenges for commercial artists to work in a capitalist system and how technological solutions like style transfer may contribute to them.

\section{Related Work}

\subsection{Style and AI}
\subsubsection{Style as a computer vision challenge}
Computer science researchers have long considered style a challenge for the systems they design, mainly through developing style transfer, the technique of extracting a style from a source image and applying it to another \cite{neuralstyletrasferreview, NSTreview}. 

Long before deep neural networks, Aaron Hertzman first proposed “image analogies” \cite{imageanalogies}, the first technique for learning generalized style transfer from a pair of transformed images and reproducing an analogous process on a new image. 
With the advances in deep learning, style transfer dramatically improved, notably with the seminal work of Gatys introduced Neural Style Transfer \cite{gatys2016image}.
Using Convolutional Neural Networks, Neural Style Transfer proved to go beyond low-level features of the image to focus on a high-level content-style disentanglement to explicitly extract style from an image and apply it to another \cite{neuralstyletrasferreview}.

With the emergence of text-to-image models like DALL-E \cite{dalle} and Stable Diffusion \cite{stable_diffusion}, it is now possible to generate images in various artistic styles by incorporating specific style keywords (e.g., cartoon, impressionist, abstract) into the prompt \cite{liu2022design}. 
These keywords guide the model to emulate the visual style of images that resemble those in its training corpus. 
Models like InST \cite{Zhang_2023_CVPR} guide text-to-image models by converting a painting into textual embeddings using an attention-based textual inversion method, enabling the creation of images with specific artistic characteristics derived from the original artwork.

Furthermore, with the input of referenced images in a target style, style tuning can be achieved by finetuning text-to-image models with techniques such as StyleDrop and LoRA. 
StyleDrop \cite{sohn2023styledrop} allows for the adaptation of specific styles in text-to-image models by fine-tuning less than 1\% of the model’s parameters, specifically targeting those that define styles such as textures and colors. 
LoRA (Low-Rank Adaptation)\cite{lora} utilizes low-rank matrices to modify weights in existing layers, enabling targeted changes with minimal parameter alterations. 
This method efficiently allows the model to adapt to new styles while preserving its original capabilities.

Regardless of the system used for style transfer, evaluating its success remains a complex issue \cite{wang2021evaluate,Karayev_2014}. 
A problem we suggest can be dissolved by enriching our understanding of style as a boundary object.

\subsubsection{HCI research on style}
To our knowledge, no HCI study paper has explicitly focused on artists’ perceptions of style transfer results. 
We review work on system design, both for and against style mimicry, that has included surveys on artists’ perceptions and studies of user interactions with generative AI models in general.

A series of HCI system works have focused on helping people apply text-to-image models using style keywords \cite{opal,promptify,dreamsheets,promptcharm}.
Several works have also developed tools to support style transfer in domains such as fashion \cite{styleme} and vector graphics generation \cite{warner2023interactive}. 

While style transfer is here to stay and has proven useful to some users, it also had a negative impact on artists. 
A survey of the text-to-image community reveals that \cite{promptingpractice} artist names are the main specifier used alongside the main subject.
This has caused much tension between artists and AI companies \cite{artistsname},  who in some cases banned the use of living artists’ names as prompts \cite{stablediffusionname}. 
Furthermore, in \cite{biasesgenai}, Srinivasan et al. show that modeling artists' styles using correlation statistics often leads to stereotyping the nature of their process and the quality of their work, leading to a biased representation of what style is.

Regarding style transfer, the infamous case of a Reddit user who fine-tuned Stable Diffusion to copy illustrator Holly Mengert’s work \cite{hollymengert} became a clear example of how illustrators’ styles might be at risk of being simply “grabbed” online to be reproduced by anyone.
In response to such threats, Shan et al.
\cite{shan2023glaze} have developed Glaze, a technique to “cloak” artists’ styles that while imperceptible by the human eye, misleads AI system trying to copy them. 
Similarly, Liang et al. \cite{adversarialpaintingstyle} introduced the AdvDM algorithm, which crafts adversarial examples that disrupt the ability of diffusion models to capture and imitate specific artistic features. 
Meanwhile, Nightshade \cite{nightshade}
proposes a prompt-specific poisoning strategy, designed to safeguard intellectual properties by subtly destabilizing text-to-image generative models, thus limiting their ability to accurately respond to targeted prompts or rendering them ineffective.

In the broader study of Text-To-Image (TTI) models in HCI, we follow researchers who foregrounded artists’ perspectives in examining text-to-image models. 
Rajcic et al. \cite{rajcic2024towards} have explored artists’s perception of TTI models through a similar methodology of using a model fine-tuned to their work. Shelby et al. \cite{shelby2024generative} have paid attention to artists' point of view about the use, harm, and harm reduction of TTI models. 
In the context of text generation, Gero et al \cite{gero2023social} show how to situate generative models within the larger socio-technical context of creation from the writers' perspective.

\subsubsection{Style as an underdefined multifaceted object}
Whether in system or study papers, we identify a common gap: a lack of a definition of style. 
In the research on style transfer, style has referred to a diverse set of objects, from artistic movements (“impressionism”), individuals’ work (usually “Van Gogh”), medium (“watercolor”), or general artistic practices (“illustration”). 
This lack of specificity and the diversity of objects referenced by the term has made researchers rely on an axiomatic “we know it when we see it” perspective of style. 

A notable exception to this gap is Elgammal's work \cite{elgammal2018shape} on classifying artistic styles, which mobilizes the theories of art historian Heinrich Wölfflin (1864-1945) \cite{wolfflin1950principles}. 
The latter provided the cultural precedent for contemporary content-style disentanglement tasks by suggesting that art history should be concerned with a formal analysis of artworks based on stylistic features (while ignoring subjects). 
Yet, this approach to artwork is one among many, and as we will show, it does not align with illustrators’ conceptions.

\section{Methodology}

\subsection{Finetuning text-to-image models}

\begin{figure*}
\centering
\includegraphics[width=0.85\textwidth]{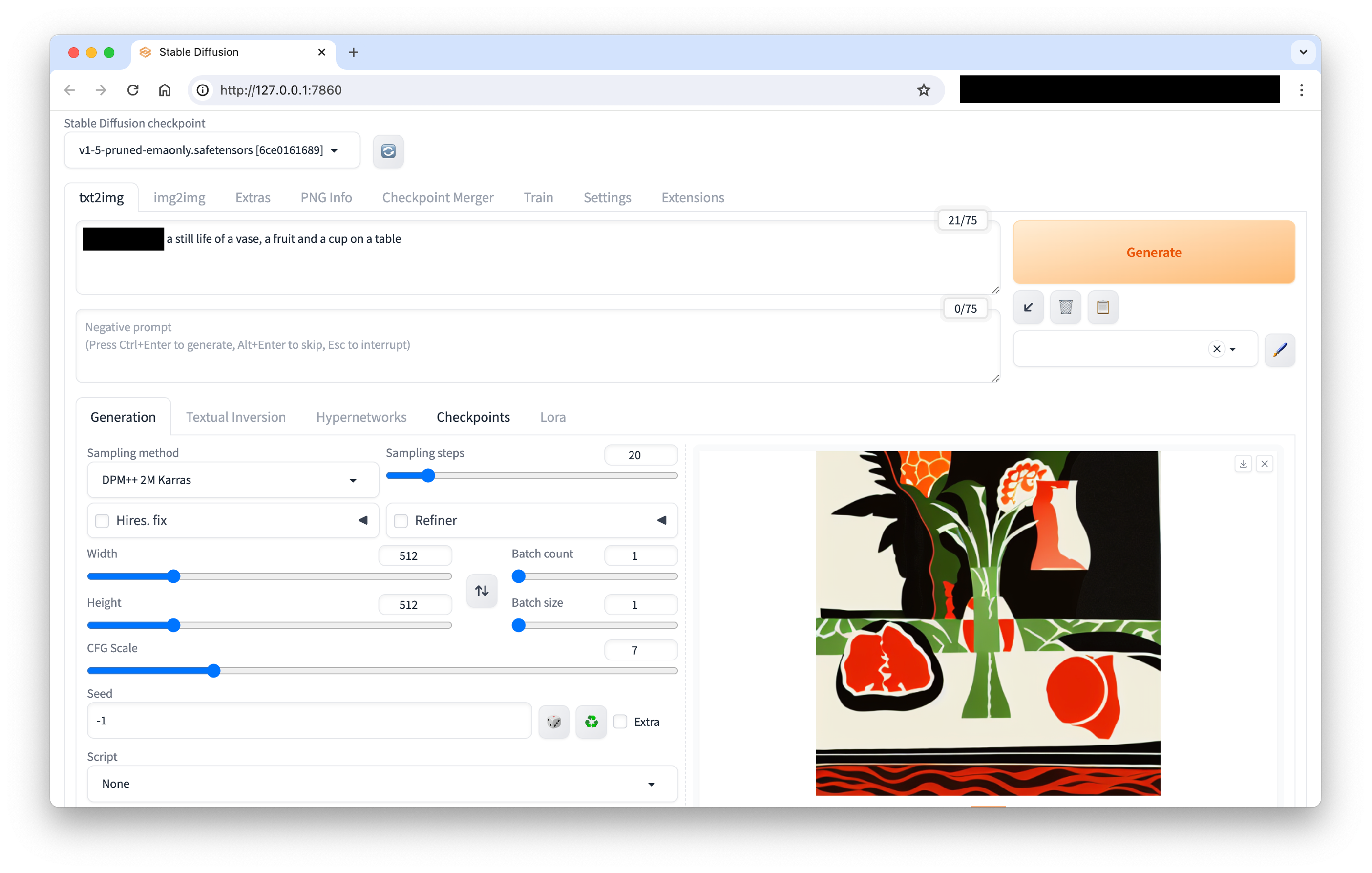}
\caption{Stable Diffusion Web User Interface used by participants}
\label{fig:webui}
\Description{Screenshot of the Stable Diffusion Web User Interface the participants used for the study. The page consists of a prompt bar in which is written the prompt “a still life of a vase, a fruit and a cup on a table”; below it a negative prompt bar, and to their right, a big orange “Generate” button. Below the prompt text bars are a series of sliders for image width, height, batch count, batch size, and sampling steps. To the right of these sliders, under the orange button, is the generated image of a flower vase on a table with some fruits.}
\end{figure*}

Our study employed Stable Diffusion v1.5 as our base model to personalize text-to-image models. 
To accomplish this, we curated a dataset consisting of 30 images from each participating illustrator's online portfolio, which were obtained with their consent. 
Each image was automatically captioned using BLIP \cite{blip}. 
For the finetuning process, we utilized LoRA \cite{lora}, which is a popular technique known for its ability to adapt models efficiently with minimal additional parameters. 
The finetuning process executes a total of 3000 training steps with a learning rate of 5e-4 for each participant’s dataset. 
The training was conducted using Colab Pro GPU, and models were stored locally in a secure and password-encrypted device.
Initial validation of the training workflow was performed using data from the first author, who is also a proficient illustrator. 
Each participant’s model was then made accessible for experimentation via the stable diffusion web interface\footnote{\url{https://github.com/AUTOMATIC1111/stable-diffusion-webui}}(Figure \ref{fig:webui}).

\subsection{Participants and study procedure}
\subsubsection{Participants}

\begin{table*} 
    \centering 
    \begin{tabular}{|l|l|l|l|l|l|}
         \hline
         \textbf{Participant}&  \textbf{Age}&  \textbf{Gender}&  \textbf{Year of experience}&  \textbf{Main clients}\\
         \hline
         P1&  32&  female&  10&  advertising, branding, editorial\\
         \hline
         P2&  32&  female&  10&  advertising, editorial, publishing\\
         \hline
         P3&  33&  male&  11&  editorial\\
         \hline
         P4&  29&  male&  8&  advertising, branding, editorial\\
         \hline
    \end{tabular}
    \caption{Participant background}
    \label{tab:parciaipants}
    
\end{table*}

We recruited four artists (2 females and 2 males, background see Table \ref{tab:parciaipants}), all professional illustrators working in New York City at the moment of the study. 
New York City is a nodal point for the illustration industry, with many publications and agencies located there. The city, with its high cost of living, is also a place where illustrators are particularly aware of the political economy of their craft, which made their insights on the matter all the more vivid.
Participants ranged from 29 to 33 years old, and they all had been working as illustrators between 8 and 11 years. 
Each illustrator has worked with editorial, advertising, and/or branding clients in the US and internationally. 
The recruitment was conducted through contacts established by one of the authors during his long-term fieldwork in the illustration industry.
Our main eligibility criteria was working professionally as an illustrator, part-time or full-time. 
This was an essential choice as we understand style holistically, contributing to artistic and economic value. 
Therefore, the insights of these professional illustrators on the value of their style in both contexts were crucial. 
This choice also contextualizes artistic practice (and style) in broader social systems instead of upholding the cultural narrative of art's exceptionalism \cite{becker2023art}.

Participants were also selected based on the diversity of their styles and mediums to enrich the diversity of perception we are looking to make salient.
None of the participants had extensive experience with generative AI and were generally curious about the perspective of having a model trained on their work. 
It is important to note that this was not the case for all the illustrators approached for this study.

Explicit consent was obtained after conversations with the participants over months prior to the study.
Embedded in the social network of these illustrators for a year of ethnographic fieldwork, the first author discussed the idea of this study first informally during conversations about AI and illustration, then more formally with each participant, thoroughly explaining the study procedure as well as the ethical measures taken to protect them. 
As we have discussed, illustrators' relationship to AI is one of dispossession, their work being fed to systems without their consent. 
Seven illustrators were approached, and three refused. One of them refused because they were not interested in the topic, and two because they felt that seeing their lifework reproduced easily by a machine would be too confronting. 
Refusal to participate in a study is a social position towards technological systems that is rich in insights, articulating, for example, a fear of exploitation~\cite{grigis2024playwriting}. 
In this context, the choice of having such a small sample has allowed us to develop trusting relationships with each participant. 
These relationships are not only about obtaining consent from participants, but also creating accountability for researchers. 
We return to the topic of participant selection in 
section \ref{sec:limitations}.

\subsubsection{Study procedure}
Participants were given access to their fine-tuned models via a stable-diffusion app to generate images during sessions that ranged from two to three hours. 
Participants were free to prompt the model however they wished, with the exception of a ``control prompt'' common to all: ``the last supper'' (Figure \ref{fig:last_supper}).  
Semi-structured interviews took place during and after the study, adapting to the participants' reactions to the generated output in the moment and exploring their overall perception at the end. Interviews focused on their perception of the quality of the output in general and as a copy of their own style. 
They were also asked to reflect on usability and past professional experiences related to style transfer. Finally, each participant was shown other participants’ model’s output and was asked to evaluate its ability to copy their style.
Using an "emergent coding" approach, i.e., inductive coding~\cite{lazar2017research}, we conducted a thematic analysis \cite{braun2006using} of the interview transcripts using the software Atlas.ti to note recurring patterns in the participants' responses.
After the first coding pass, the team discussed the various codes to come to an agreement on the most important ones. 
The high-level categories have been used to structure the Findings section to stay as close as possible to the data.

\subsection{Note on researchers' positionality}
Considering our goal to locate the perception of style, we heed the call from multiple HCI researchers to disclose researchers’ positionality concerning their research \cite{intersectionalhci,feministhcimethodology,margins}. 
Notably, the first author is a PhD researcher in social anthropology and has worked as a professional illustrator for several years. 
The artists were recruited from participants in this author’s year-long ethnographic fieldwork in the illustration industry. 
We consider this intimate embedding in the illustration industry, both as an observer and practitioner, to be a position worth knowing for the reader. 
On the one hand, this has allowed us to access a deeper understanding of illustrators’ lives beyond the study's scope and earn their trust regarding a sensitive topic like style mimicry. 
On the other hand, this proximity to the impact of generative AI also grounds our research within a clear agenda of supporting the struggle of freelance artists.

\section{Findings}
\label{sec:findings}
Our findings show that evaluating style transfer's success, far from being a neutral assessment of a model’s performance, is a socially situated perceptual task, i.e., not everyone sees the output of style transfer the same way. 
Our questions focused on artists’ perception of (1) what was accurately reproduced of their style, (2) what made the model unable to reproduce their style, (3) the usefulness of the model in their work, and (4) their perception of other participants’ output. 
These themes will outline significant differences in the ways AI models and artists perform style.

\subsection{Texture, colors, and other successful fragments}
\label{sec:texture_colors_and_other_successful_fragments}
\label{sec:sucessful-fragments}
When the participants were asked to evaluate the model's success in copying their style, they identified isolated elements of the images, referencing low-level aspects of the images such as colors, textures, shading, and so on. 
At first, struggling to find anything similar to their work in this copy, closer inspection would lead them to identify small pieces of the images they saw as accurate depictions of their style, stating that “[…] some of these underarm shadows are pretty good” [P3] or “the lines on the shirt, I think it picked up on that. When I draw shirts, I would do folds and stuff [P4].” 
This granular evaluation of stylistic output contrasted usual evaluations of style transfer which tends to deal with “style” broadly defined, if defined at all. 
We now turn to texture, color, and shading in particular.

\subsubsection{Texture}

\begin{figure}
\centering
\includegraphics[width=0.48\textwidth]{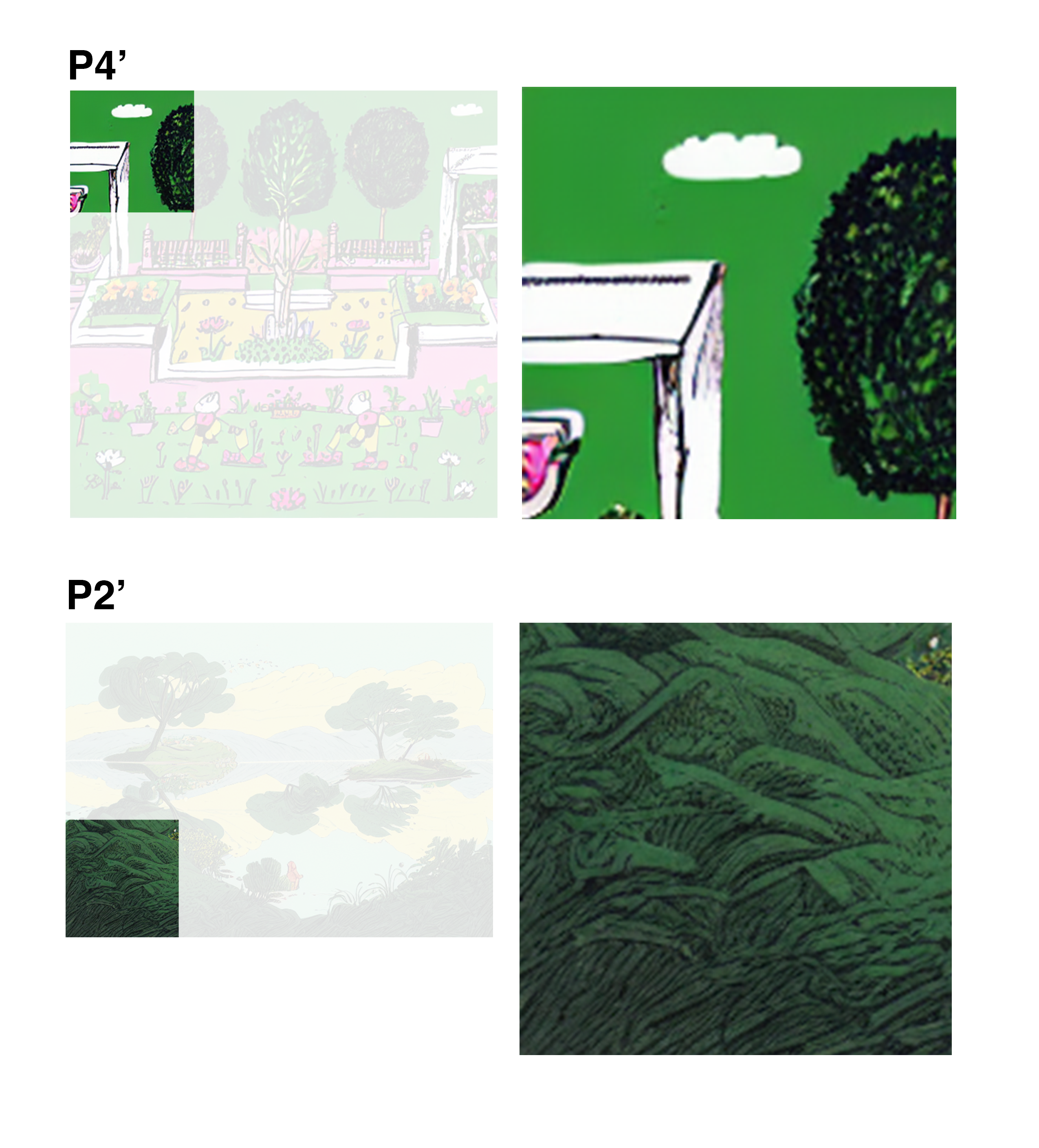}
\caption{Textures evaluated as successfully copied}
\label{fig:texture}
\Description{2 pairs of 2 images are stacked on each other; the top pair is labeled P4’, and the bottom is labeled P2’. In each pair, the first image is an AI generated image with a white, 80\% opacity coverage, leaving the image barely visible except for one small square left clear. The second image of each pair is that small clear square of the first image blown up, showing the texture of a small patch of an image.}
\end{figure}

Texture was generally accurately transferred from artists’ work to the output, for example, evaluating another artist’s [P2] output, this illustrator was impressed by its quality:
\begin{quote}
\textit{
[P1]: “I'm just really surprised by how it [the model] was able to capture the essence of these textures, I don't know.”
}
\end{quote}
For artists, texture refers to a wide variety of visual elements, from their “mark-making” techniques (e.g., cross-hatching), the grain of a pencil (e.g., 6b vs. HB pencil), or the overall texture of a paper. 
While this conception of texture overlaps that of computer science, it also points to more specific artifacts than the overarching structure of an entire image.
Participants would often refer to the texture of a particular chunk of the images (Figure \ref{fig:texture}). 
One illustrator was surprised by the accurate similarity of the clouds’ texture with his work:
\begin{quote}
\textit{
[P4]: “Oh, the clouds are interesting... I like this little texture [the model] added here. Like some of the decisions... I would do a lot... Yeah, I think it picked up on that. It's a pretty prominent feature on my work if you want to pick out like clouds.”
}
\end{quote}
Here, the texture P4 references is not a quality of the entire image but a localized element in the generation of clouds only.
Another commented on a specific texture coming from a particular image from the dataset:
\begin{quote}
\textit{
[P2]: “Well, better than anyone I just see specific little things that I feel like that might have been just taken from one of the images, and that's like where the texture comes from.”
}
\end{quote}

\subsubsection{Color}
Another aspect of the output that artists positively evaluated was color. 
Most of them were surprised and pleased by the accuracy of the color palette generated in the output, stating, “Then you see the composition and you see the colors and I think that's what it does best” [P2] and “The colors are quite successful.
That's an interesting thing” [P1]. 
Even in assessing other participants, color was an accurately copied element: “I think color wise they nailed her” says [P1], talking about what the model succeeded in reproducing about [P2]'s work.
Recognizing someone’s work by their color palette points to how important color is to style as an index of their identity and uniqueness. 

But again, when reflecting on their illustration practice, artists agreed that color serves a purpose, without which color transfer does not contribute to style transfer, as one of the artists explained:
\begin{quote}
\textit{
[P4]: “And the color choices are close, but for the garden one, for example, it is kind of far, I just wouldn't use that neon [green], like I used it in my work before, but I wouldn't pair it with... certain colors I wouldn't pair with certain ones. […]”
}
\end{quote}

\subsubsection{Shading}

\begin{figure}
\centering
\includegraphics[width=0.48\textwidth]{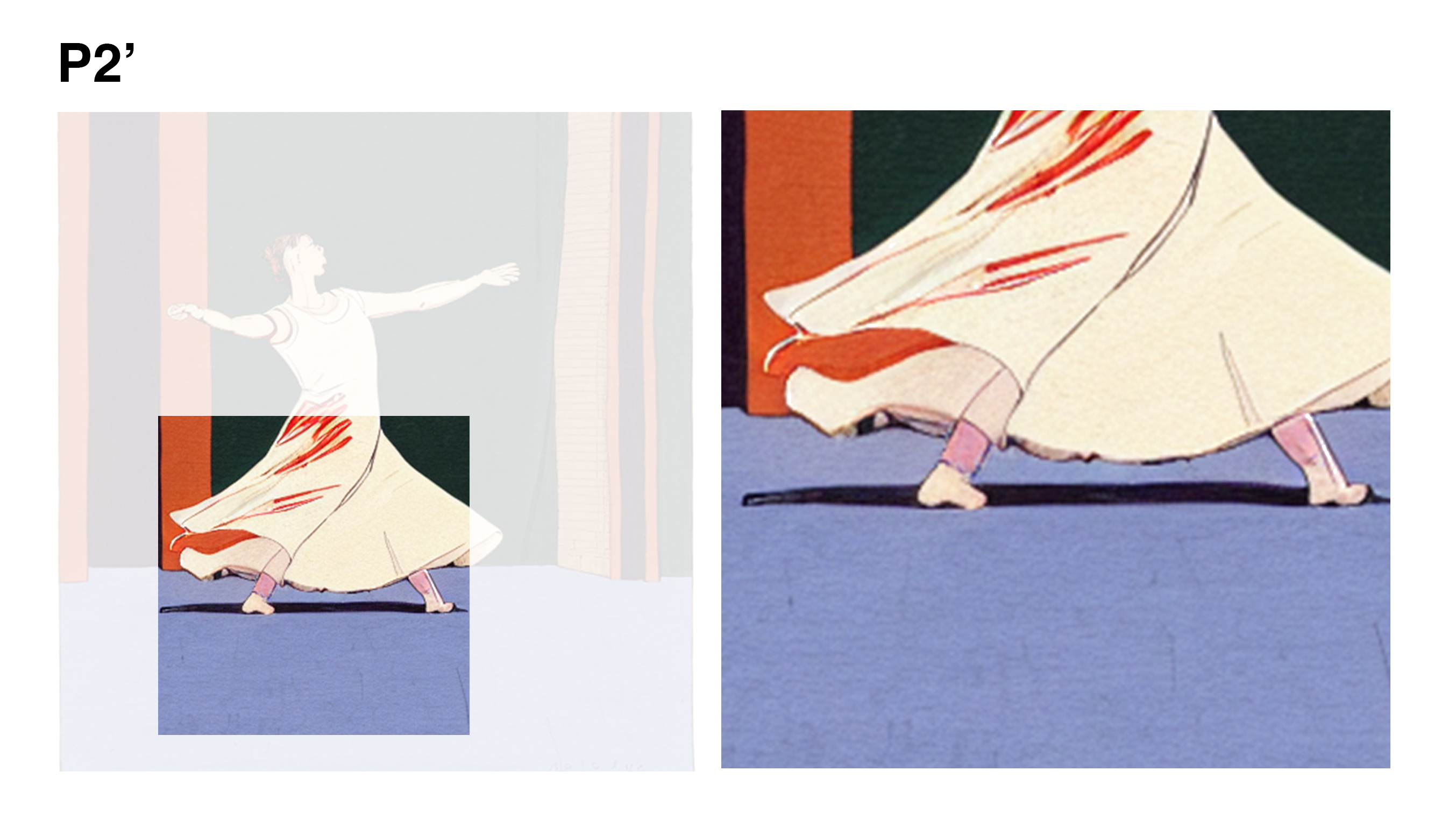}
\caption{Shading evaluated as successfully copied}
\label{fig:shading}
\Description{A pair of images labeled P2’. The first image is an AI generated image with a white, 80\% opacity coverage, leaving the image barely visible except for one small square left clear. The second image is that small clear square of the first image blown up, showing the shading under a character's dress.}
\end{figure}

Several participants reported being pleased with the shading of parts of the images:
\begin{quote}
\textit{
[P4]: “I don't know. I think it's the way that shadows are drawn and the kind of graphicness they have. It's like graphic but soft at the same time.[...] I'm actually quite... I'm impressed by this shading here [points to a door on the image].”
}
\end{quote}
\begin{quote}
\textit{
[P2]: “I mean, I can see some similarities, but it's so far off still. It's hilarious. Yeah, probably in the dress, especially the strong shadows.[...] I do that sometimes, the way it shades with a little lines, that's it.”
}
\end{quote}
Shading is obviously limited to particular elements of the images and was not always present in the output (Figure \ref{fig:shading}). 
Yet, even when the overall results did not convince them, participants often found the shading to be accurately copied.

\subsubsection{Textures, colors, and shading did not equate style}
Textures, colors, and shading are all discreet elements whose results never exceeded the sum of their parts for the artists involved. 
The model was good at reproducing such isolated, low-level elements of each dataset but only in localized places and not equally throughout every output. 
Most importantly, artists' evaluations show that however successful the model was at reproducing some parts of their style, it was only doing so in a way that never exceeded the sum of its parts:
\begin{quote}
\textit{
[P3]: “There's a gesture towards simplified colors. But it's very inelegant.”
}
\end{quote}

\begin{quote}
\textit{
[P2]: “I mean, the same way I can see me in my images, like they are compositional things, like the kind of relation of how big is a head compared to a body, the color choices and some little artifacts like mark-making, it gets right, but then it just chops them up so weirdly.”
}
\end{quote}

This “chopped-up” version of style mimicry points to a crucial difference in how artists and models perform style. 
Interestingly, our findings closely match the features Garces et al. \cite{garces2014similarity} have proposed as metrics for stylistic similarity: texture, shading, colors, and stroke. 
Yet, as we’ve shown, while the artists deemed these elements of style transfer indistinguishable from their work, it was always in a narrow and limited sense. 
In other words, while the model successfully transferred some qualities of the dataset, these could hardly be qualified to amount to the artists’ style.

\subsection{Challenging the aesthetic-content disentanglement task in style transfer}
\label{sec:content-style-disentanglement}
When asked about the model's limitations in generating an accurate copy of their style, participants often pointed to the difficulty of working with style-content disentanglement.

Style transfer is based on a “style-content disentanglement” task \cite{neuralstyletrasferreview}. 
While the semantic bias has been addressed in research focusing on the limitation of natural language as input \cite{opal}, we suggest that the problem with a separation between semantic content and aesthetics runs more profound than the issue of text vs. image. 
Our finding indicates that artists understand style as pertaining both to aesthetic treatment and semantic content and moving seamlessly from one level to the other, making style inherently an entanglement of both.

\subsubsection{Texture without content}
Returning to texture for an instant, its positive evaluation was limited by how it was used and linked to semantic content. 
For artists, texture has a purpose and is linked to intentional use. 
As one participant explains, it can not be applied indiscriminately on every pixel of an image; it must make sense in relationship to the content of the image:
\begin{quote}
\textit{
[P2]: “I like the texture. But like what is it? Is it grass? Or is it rocks? Because I would use a sort of texture for rocks. And I would use a sort of texture for grass. And this is a mixture of both. And I wouldn't draw it that way. But I kind of get how it's taken from my work. And I like it. I mean, I could see myself doing something similar or having done something similar.”
}
\end{quote}

Texture is guided by the object represented in the image and the effect the artist aims to achieve.
Another artist distinguished between texture and the “act of drawing” it is supposed to index:
\begin{quote}
\textit{
[P3]: “The biggest [flaw] to me [is that it] feels like the actual act of drawing is not present, and that's like some textural feeling but also [it] has to do with how an image is constructed.”
}
\end{quote}
This “textural feeling” only makes sense with how an “image is constructed”, indicating that for texture to contribute to style, it needs to be applied based on the semantic logic of the image. For artists, an image's texture, rather than a superficial layer, is an integral part of meaning-making.

\subsubsection{{Semantic objects are stylistically important}}

\begin{figure*}
\centering
\includegraphics[width=0.9\textwidth]{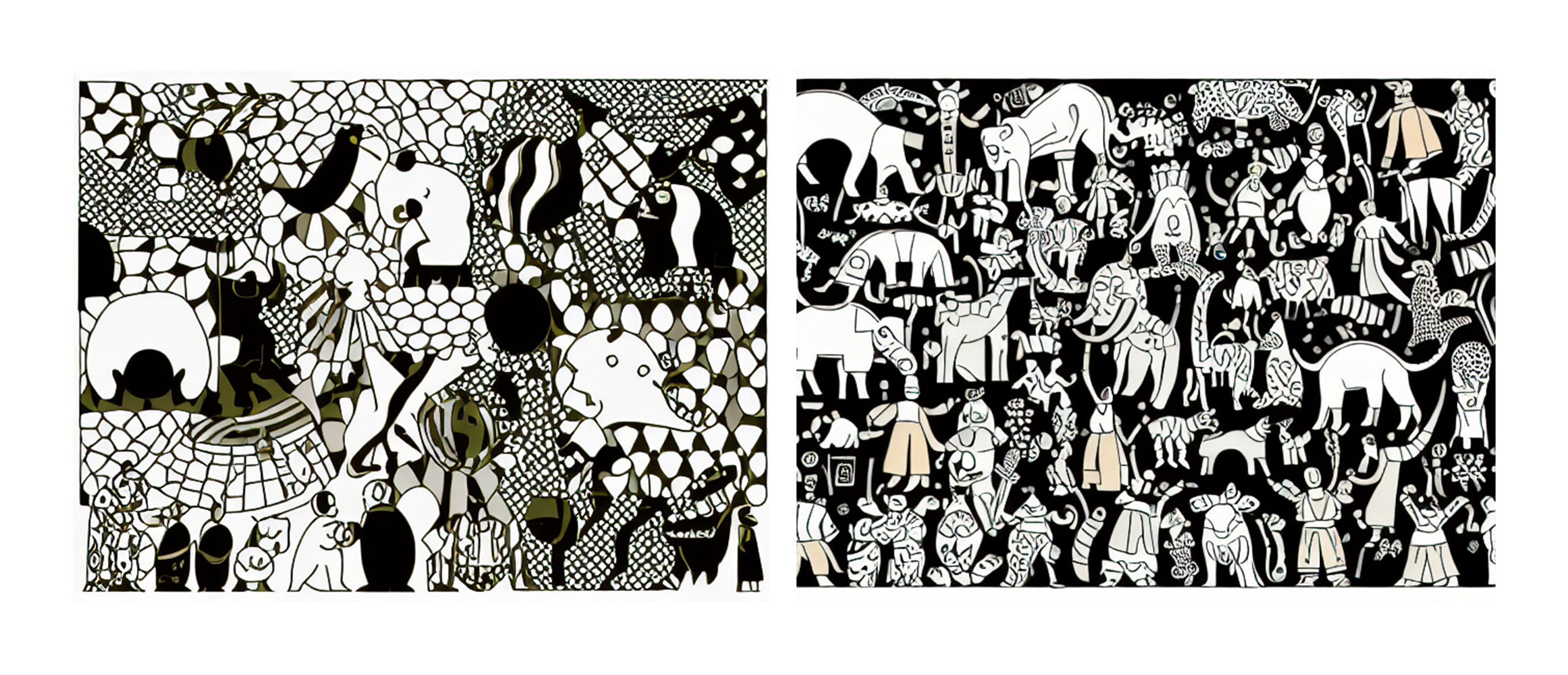}
\caption{Two style transfer outputs from the prompt "Zoo animals and circus" by [P2]}
\label{fig:image4}
\Description{Two illustrations side by side. The illustrations were generated by P2 using our style transfer model with the prompt "Zoo animals and circus."}
\end{figure*}

Each artist has semantic elements they always draw, or conversely never do, the presence and absence of which become an essential part of their style over time. Many participants prompted the models to generate subjects they generally never draw, and therefore absent from the dataset. In such instances, the model struggled to generate output that was interesting to them.

As this illustrator explains:
\begin{quote}
\textit{
[P1]: “ Zoo with animals and circus... I think if someone asked me to do a zoo I would literally convulse and say No. I don't think I could do it. it's gonna be so crazy, I don't even know what [the model] is going to come out with.”
}
\end{quote}
[P1] expresses here the difficulty of picturing her style, including the semantic elements of animals.
Thus the result of the prompt “a zoo and a circus” generated an output considered an outlier compared to the more accurate copies that participants generated (Figure \ref{fig:image4}). 
She considered the result not to be representative of her style and joked that she too would actually be bad at drawing animals: 
\begin{quote}
\textit{
[P1]: “But it's so funny. I don't have any animals that [the model] was like ‘but I know what an animal looks like, how would [P1] draw one?’ and I would draw an animal like really scary just like this.[laughs].”
}
\end{quote}

Looking at the output of her model, another illustrator explains why semantic content and aesthetic choices are completely inseparable in “conveying her style”:
\begin{quote}
\textit{
[P2]: “I feel like to properly copy someone's style, the subject of the content of illustration has to change. I remember we had to do self-portraits once for our exhibition at school, and I drew myself holding myself in my hands. And of course, it didn't really fit the format because everyone else had a face. [But] I don't like to just draw face, I don't know how to convey style in just the face.”
}
\end{quote}
As [P2] expresses here, even within such a semantically normative genre as the self-portrait, what made her style hers was the alteration of the "subject of the content of the illustration" alongside the aesthetic aspect of it.  

\begin{figure*}
\centering
\includegraphics[width=0.9\textwidth]{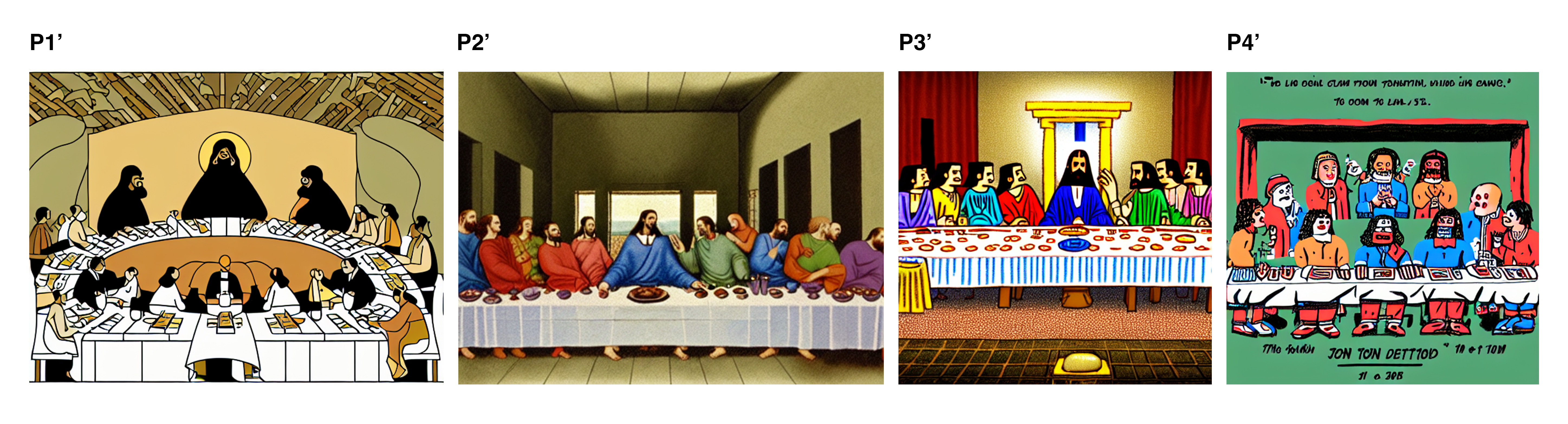}
\caption{Images generated with the control prompt "the last supper"}
\label{fig:last_supper}
\Description{A series of 4 illustrations side by side, each labeled from P1’ to P4’. They all depict the Last Supper in what the model has interpreted as the style of each participant.}
\end{figure*}

Prior to the interviews, we generated some images with each participant's model using generic prompts such as "two persons talking" or the control prompt "the last supper" (see Figure \ref{fig:last_supper}). Some participants found that the model reproduced their style less accurately with these prompts than their own. For example, some of P2's prompts were: \begin{quote}
\textit{"a tree and and over proportionally big flower on an island in the foreground with a person sleeping underneath it. in the background a lake and mountain. wind is blowing through the tree. it is dusk. birds are flying."}
\end{quote}
or 
\begin{quote}
\textit{"a man is standing in an empty street, reaching up to touch a bird is flying above him." }
\end{quote}
Similarly, P3 spent a considerable amount of time prompting an image of a character he enjoys drawing, Humpty Dumpty, iterating numerous times to give the right stylistic cues:
\begin{quote}
\textit{
"An egg person is sitting on a brick wall. He is an egg with a face and short arms and legs. His hands are very big. His world is not three dimensional. He has a big blank smile. His eyes are blank. He wears a bowtie. His clothes are simple, they do not have intricate features. There is an apple sitting next to him on the wall."}
\end{quote}
These prompts, unlike our control ones, are rich with idiosyncratic semantic elements (P2's birds, over proportionate flowers, or P3's Humpty Dumpty with a bowtie) and stylistic directions, which are intrinsic to their evaluation of style transfer as accurate. Commenting on the difference between the output generated by our respective prompts, P2 explains:
\begin{quote}
\textit{
[P2]: ``It takes me a lot of time and effort to see it [the similarity]. Especially with these images that you generated. The other ones [that she generated during the experiment] that was quicker to see the resemblances because those are also things that are different from my work and different from how I approach drawing.''}
\end{quote}
Evaluating the ability of the model to reproduce style was highly dependent on the semantic elements chosen to generate output. This points once again to the impossibility to disentangle style from content, or at least its uselessness for artists.
Semantic elements therefore contribute to style as a whole, rather than be separate from it.

\subsubsection{Aesthetic choices are semantically meaningful}
If semantics and aesthetics are always linked for illustrators, they also articulated that aesthetic choices are semantic. 
Reflecting on what makes interesting images, a participant shares how the separation between semantic and aesthetic content is not conducive of good work, which inherently spans both:
\begin{quote}
\textit{
[P1]: “we are trying to like, not draw just the scene for what it is [but] push the design, literal box, and see how to make it. […] Because a lot of illustrators, you probably know that, like, they really just draw what was being asked of them. And it's not like the most… they're not pushing, in terms of composition. […] There's some people that are really good at it, I feel like [illustrator’s name] is really good at like, playing with, like, the actual shape of the box that she has, and that adds the dynamism. But some people just don't do any of that.”
}
\end{quote}

In this excerpt, this illustrator uses the notion of “pushing the box” both literally and metaphorically. 
On the one hand, she refers to pushing the literal box, the frame of the image, and playing with composition but also thinking of what is not present in the image (“off-screen”) or even crafting illustrations with elements going outside of the frame to bleed into the page. 
On the other hand, by contrasting illustrators who “just draw what was being asked of them”, pushing the box also means having a reflexive practice that exceeds and transcends the division between semantic content and aesthetic treatment. 
While some illustrators take a brief and draw it literally, this illustrator suggests illustration is about doing more and creating a complex relationship between what is shown and how it is shown. 
Giving an example of an illustrator’s work, she describes how aesthetic choices are conceptual, even when impossible to describe:
\begin{quote}
\textit{
[P1]: “I think maybe a concept is, when it's like distorting an image, you take objects that are common that [we] are able to understand, but you like rearrange them in a way that's trying to say something bigger than itself.”
}
\end{quote}
For her, the aesthetic manipulation of the image, “distorting” is the way to “say something”. This entanglement of aesthetics and semantics is therefore crucial to illustrators’ style. 
Our findings indicate that illustrators rely on both semantic and aesthetic elements to convey their style in their work, making the goal of disentangling content from aesthetics useless at best and damaging at worst. As we will show in the discussion, while this conception of style transfer does not benefit artists, it does benefit other stakeholders of the illustration industry.

\subsection{Style as an emergent quality of creative work}
\label{sec:style_as_an_emergent_quality_of_creative_work}
As we have shown so far, our interviews reveal that style is a multifaceted concept encompassing textures, colors, shadowing, mark-making, concepts, subjects, identities, and more. 
Closing the interviews, we asked the participants whether they worried about their potential replacement by models who could copy their styles. 
They all shared that they did not feel worried at all.  

This answer may be surprising considering how much fearmongering and “hype” generative models have generated in the past few years. 
One might argue that the quality of the output of our fine-tuning models was the cause of their reservations, but looking into their answers points to a more complex reason:
\begin{quote}
\textit{
[P4]: “[...]if one day people are evil and they take my images and directly feed it like that [to a model]. Of course, I would hate that. But I'm like, you know,[…] in the 1980s, you will have to get a disc to get 10 fonts on your computer, today, we literally have every resource we can like in our lives. And people are still making like terrible design choices. They literally make the worst stuff ever. \textbf{I just don't think that's about the resources we have. I think it's about taste and intention.} It doesn't matter if AI exists or not in those domains, if someone already has great taste, then they probably will know well enough to not use AI.”
}
\end{quote}
In this quote, the illustrator outlines what is inherently missing from style transfer, generative models in general, is taste. 
The topic of taste and generative AI exceeds the scope of this paper (for a survey of the topic, see \cite{hullman2023artificialintelligenceaestheticjudgment}), but we address it here as a more general form of aesthetic evaluation.
Comparing his predicament with the changes in typeface availability since the 1980s, the illustrator addresses pre-emptively arguments suggesting that quantity will beget quality (larger dataset, bigger compute, etc.) to argue that without taste, style is useless. 
By centering taste as a key element of illustration work, he gives the key to the shortcomings raised so far in style transfer. 
Taste ties texture, colors, and other low-level aesthetic elements to their context and ties the semantic aspect of an image with the aesthetic choices made to depict it. 
By performing a version of style that relies on its crystallization, extraction, and application regardless of context, generative models preclude the possibility for artists to use them in meaningful ways.

Another illustrator, discussing the potential use of the model in her workflow, articulates the way she works, the role of taste and experimentation in her style:
\begin{quote}
\textit{
[P2]: “Whenever I get a brief or something I've never done before, and that doesn't feel quite comfortable […] I have to grow. And I have to expand my knowledge of what's possible. And I maintain my taste and sensibility, but I have to expand that. There's some sort of pain of growth attached to that, it's difficult. Every time I have to do that I’m a little bit worried, and I don't like my first sketches,\textbf{ but then I discover something new, and my work changes. And [the model] can't do that.} It needs me to feed it the new images, so it knows how to do it. So in a way, it's not very usable, because you can only reproduce what has already been done.”
}
\end{quote}
Evident in this citation is the way style, rather than being a fixed cognitive plan to follow and apply in every new image, is an emergent quality of creative practice \cite{Suchman1987PlansAS}, negotiated in the activity of drawing. 
Here, rather than being characterized by similarity, it is difference that makes style such a powerful tool for artists. 
Note that the criticism of the model does not rely on its inability to reproduce style consistently but its inability to deviate from it while still being guided by the same taste and sensibility.

\subsection{Differential perceptions of the output}
\label{sec:differential_perceptions_of_outputs}
So far, we have seen how the perspective of illustrators has enabled us to specify exactly what is successfully transferred or not by style transfer. 
Yet, our interviews have shown that different participants disagreed on what they saw even when looking at the same output images. 
The evaluation of style transfer success was not homogenous. 
For all participants, a major  takeaway of the experiment was an appreciation for these other “ways of seeing” \cite{berger2008ways} their own work:  
\begin{quote}
\textit{
[P1]: “It's really interesting to think about how this tool is reshaping what we think about or making us confront how we see ourselves, how other people see us and what different people pick up, or attribute value to.”
}
\end{quote}

During and after the interviews, some participants went on their website (where the dataset was pulled from) to remove some images they felt conveyed an outdated version of their style. 
Others refused to hear what other participants thought of the output of their model, not wanting to be aware of others' perceptions. 

Crucially, we found that the perception of the model's success was aligned with the participant’s proximity to the style evaluated, i.e., how familiar they were to it. 
While, as we have shown, none of the participants felt that the model was really successful in copying their own style, they were also more likely to find other participants’ output to look like their work:
\begin{quote}
\textit{
[P1 about P3]: “This looks like yeah, it's... they got him.”
}
\end{quote}

 \begin{quote}
\textit{
[P1 about P2]: “Well, these are so scary. I think color wise they nailed her.”
}
\end{quote}

 \begin{quote}
\textit{
[P3 about P4]: “I mean, like, it's not a good composition it's not communicating anything. But these characters could live in [P4]’s world.”
}
\end{quote}

 \begin{quote}
\textit{
[P4 about P2]: “But then, like when I see [P2] it's like, oh, this is similar. But when I see mine, I'm like [grimace] because I'm trying to judge it against my work. And I know my work best because it's on my mind rather from... I already have a database in my head that is the most accurate representation of my work. I don't know, I just don't feel threatened.”
}
\end{quote}

As this last participant explains, proximity to someone’s images, and ultimately their own, makes for a more refined, granular appreciation of (and likely biased toward) their quality and therefore of the success of a copy. 
In line with this argument, we saw that participants who were the closest (some of them close friends) were more likely to pick up on small differences in each other’s work than with others: [P1]: “This boot is very not [P3]! [showing the buckle shoe].” 

Consider this example of two participants, P1 and P2, evaluating the output of P2. 
The first one, reflecting on the limitation of the output and what she deems central to her style, points to her favorite aspect of her drawing, which she sees missing from the output:
\begin{quote}
\textit{
[P2]: “[…]how much joy I take in drawing a single line, how I spend time with it makes some of them thicker, and some of them thinner, how I tried to balance things out”
}
\end{quote}

Conversely, the other participant, looking at the same images, found the results surprisingly accurate, down to the exact aspects of mark-making and brush strokes the first participant deemed missing:
\begin{quote}
\textit{
[P1]: “I'm really, really messed up by the fact that they were able… I don't really know her work that well, but I feel like... […] or even like something in the way she uses brushes, and they're not always like, they have a little extra but sometimes they are very clean and it's like, [the model] even do that.”
}
\end{quote}

As a disclaimer for the quality of her evaluation, P1 states she does not know P2’s work very well, pointing to how a stronger familiarity with it might change her perspective. These findings demonstrate that the perception and the evaluation of style are not only unique to different evaluators, they depend on their proximity to the style. 
One’s successful style transfer is another’s failure, and even a narrow category like “illustrator” includes a variety of perspectives that do not necessarily align. 
Evaluating style transfer from the “view of nowhere” \cite{haraway} is therefore bound to be limiting. 
In our discussion, we explore how these differences of perception unfold in the work of illustrators with their clients.

\subsection{The political economy of style transfer
}
As our findings suggest so far, style transfer generates a sort of “shallow copy” of style, which different actors can see differently as successful or not.
We now turn to how illustrators linked what was generated by the model to how their clients relate to style.

Participants often made digressions about their professional work during our interviews and some artists have likened the model’s performance of style to the already existing appropriation of artists’ work in the industry:
\begin{quote}
\textit{
[P2]: “\textbf{The way they [clients] try to grab illustration, it has always been grabbed. And it's always been grabbed shittily.} Like, there's a reason why we still exist. Yeah, because if you could just grab it like that, then people will just do that. It's just not quite as possible.”
}
\end{quote}

This “grab” of illustration (a striking term echoing the larger ``data grab'' of tech companies \cite{mejias2024data}) by clients as we are discussing style transfer models points to how both practices share a similar logic. 
This match is evidenced by how some clients, constantly seeking to optimize efficiency and lower costs, increasingly use generative AI as a blueprint for how to work with illustrators:
\begin{quote}
\textit{Q: “Do you think this [style transfer] is enough for clients?” \newline
[P2]: “I think they might think that because \textbf{I do get these briefs of people like this, the client who wants me to do like 30 images in two weeks plus animation, I think they would love to just generate something} and they told me ‘We like these [elements] of your images, we want you to draw it […]’. You want like a nice image and a specific style, \textbf{then kind of prompting me like an AI,} but they don't quite understand that some of the prompts they are giving me aren't possible to implement with that same exact same language. They gave me another image of a city that I did for the [client name] project, that's what they liked, but then they tell me ‘now do that about this close-up of a train’. And it's a different image like I have to work with it differently and I'm going to do differently to get that same kind of feeling that you want, I have to find new ways to do it. \textbf{And this AI is very much showing me it can't do that, it can't apply the same language to a different image. That's what it promises but it's not working.}”
}
\end{quote}
For this participant, the presence of generative models has incentivized clients to interact with artists as they would with models and to require the same efficiency from them. They “prompt [artists] like an AI,” not only creating unrealistic work conditions (30 images in two weeks) but also objectifying their work in fragmented, decontextualized ways (“we like these elements of your images”). 

Yet, for all the reasons outlined in our findings, the way illustrators create style is not comparable to what the models can generate. 
For them, style is about process and structure, which aesthetics follow. While clients might be tempted to conflate an artist's work with an AI's output, the same illustrator explained:
\begin{quote}
\textit{
[P2]: “they're also dependent on me not being an AI. Yeah, because they want the same feeling for different subjects, but that means working very differently with the subjects to make them feel the way they wanted to feel.”
}
\end{quote}
Since the release of generative AI models, many professional artists have reported a loss of income, with clients eager to switch to cheaper options \cite{lossincome}. 
While this displacement and replacement of stylistic labor by AI solutions seemingly points to an equivalence between the two, our study shows this is not true. 
What these models generate in terms of style is not similar to either the result or process of artists’ work.

\section{Discussion}
On the one hand, style transfer research suggests that AI models are able to generate images in the style of illustrators, implying that there is some sort of equivalence between the two. On the other hand, our findings show that illustrators, when given the opportunity to use these models, are unimpressed by its output. How can we make sense of this tension?
In this section, we discuss this paradox by analyzing the social context of style transfer, and particularly its political economy.

Connecting their perception of style transfer’s results and limitations to the challenges they already face in their industry (outlined in section \ref{sec:political_economy_style}), we discuss our findings in the context of the larger political economy of illustration as a profession, which constitutes illustrators’ reality. 
First, we introduce the concept of ``boundary objects'' to understand how style, rather than a single ``thing'' to be extracted from the data, is a complex multifaceted object seen and mobilized differently by various communities.
We then examine how style transfer creates a partial analogy to artists' style. Finally we show style transfer can be seen as a supply-chain optimization tool rather than merely a creativity support tool. 

\subsection{Style as boundary object}
Style sits at the intersection of many stakeholders (artists, their clients, managers, agents, etc.) who desire to do different things with images.  
Their different, sometimes competing, goals call for an analysis that can accommodate a multiplicity of views of the same object. To do this, we propose conceptualizing style as a “boundary object.”

In Star and Griesemer’s words \cite{star1989institutional}, “boundary objects are[...] both plastic enough to adapt to local needs and constraints of the several parties employing them, yet robust enough to maintain a common identity across sites.” 
They have the unique quality of being both generally recognizable as one thing by various communities yet may have specific uses in each one.
Therefore, boundary objects allow communication between social worlds (e.g., computer science and creative industry). 

Understanding style as a boundary object enables us to address the plurality of style perceptions to question the validity of universalizing metrics to evaluate stylistic similarity. 
This means considering the designers and users of style transfer as partially interacting with style from different perspectives \cite{translation,boundaryobject2,tangiblesocial}. Engaging with the critical literature that has questioned the universality of datasets \cite{crawford2021excavating,datasetpolitics,genealogydatasets}, output \cite{wevers2018unmasking}, or the positionality of data workers  \cite{dataproduction,annotators}, we use boundary objects to show that what tasks like style transfer generate also embed a standpoint \cite{feministhci}\footnote{While our aim here is to tend to the social plurality of style, we acknowledge there is an intuitive common perception of it grounded in cognition\cite{augustin2008style, cupchik2009viewing}.}.

Boundary objects also raise our awareness of power dynamics in the generation, perception, and control of styles. 
By facilitating or hindering collaboration between different communities, they can contribute to the power imbalance between them \cite{artinterpower}.
In the case of the style generated by style transfer, we will show that it hinders collaboration with artists but facilitates the capitalist extraction of value from their work by their clients. 
In doing so, we link style transfer to the political economy of illustration work, responding to feminist and STS scholars' call to connect technological solutions to their political context and impact \cite{winner2017artifacts,suchman2002located,rogaway2015moral}.

The semiotics of style are complex and worth a note to clarify what we talk about when we talk about style as a boundary object. 
Style can refer both to an abstract idea, a ``generative principle'' \cite{wilf2013media} that guides the production of artworks; and the concrete occurrences, the artworks that embody this idea in the visible world. 
In Charles S. Peirce \cite{peirce1974collected} terminology, these levels are known as the distinction between two levels of signs, type and token. 
Types are ideal and abstract, for example, the idea of the letter ``A.'' 
Tokens, on the other hand, are the concrete occurrences we encounter in the world, e.g., the lowercase ``a,'' the capitalized ``A,'' or a hand-drawn ``a.'' 
Thus various tokens can point to the same type. 
Both are, in practice, inseparable, as the type can only exist, socially, through its tokens \cite{sep-types-tokens}, e.g., we only get a sense of ``a style'' through the perception of images conveying it\footnote
{For an analysis of style and computation in semiotic terms, see \cite{wilf2013media}.}. 
For our analysis, this entanglement between abstract types and concrete tokens is important as it is exactly because of this that style can function as a boundary object. 
It can function both as an abstract common object to various communities but in its concrete, localized actualization, it differs widely, creating tensions and synergies.

\subsection{A partial version of style}

As Langdon Winner \cite{winner2017artifacts} pointed out, the politics of technology do not only reside in its use but also in its very design.
While many style transfer papers mention the potential misuse of their system, we suggest that the design of a style transfer task inherently fits the objectifying strategies of corporations seeking to benefit from artistic labor, regardless of its use.

As evidenced by the findings in section~\ref{sec:differential_perceptions_of_outputs}, the evaluation of the success of style transfer is socially situated, with ingroup differences of perception noticeable even in such a small sample as ours. 
Moving away from approaches that would dismiss such differences as self-serving biases, we suggest that these findings points to how style, as a boundary object, can be seen differently by various actors, while still being understood as common to all. 
In this study, illustrators have not found the model to be useful for a variety of reasons: style transfer only generates limited similarities with artists' work (section~\ref{sec:texture_colors_and_other_successful_fragments}), is limited to the disentanglement of style and content (section~\ref{sec:content-style-disentanglement}), and creates a fixed snapshot of aesthetics rather than enables the emergence of style within creative practice (section~\ref{sec:style_as_an_emergent_quality_of_creative_work}).
These limitations point to how, rather than a single object that is either style or not, AI-generated styles are but one version of what style can be. 
It is only one facet of a larger, more complex object.

Note that this nuanced understanding of computation styles was present in the early days of style transfer.
For example, this echoes a distinction that Aaron Hertzmann introduced in his seminal work on “image analogies” \cite{imageanalogies}. 
In biology, analogous elements share functional similarities but no structural kinship (like the wings of a bat and those of a bee). 
Homologous elements, on the other hand, might look different but share structural similarities that point to common ancestors (the human hand and the bat wing, for example) \cite{wilkins2014homology}.
What style transfer generates is, for some stakeholders, in some contexts, analogous to artists’ style, but only on the surface. 
Unfortunately, this careful nuance between analogies and homologies has since been lost in discussions of style transfer, which led to hyperbolic statements that artists’ styles and models’ results are equivalent. 

To understand how such a partial version of style can seemingly have such a disruptive impact on the illustration industry, we must move away from style as a mere aesthetic problem and return to the political economy of illustration, and who profits from the extraction of style from images, if not artists.

\subsection{From creativity support to supply chain optimization tool}

Artists have always been an unruly class of workers under capitalism, their work unwieldy for the predictable, risk-averse needs of capitalist production \cite{bedard2020art,child2019working,beech}. 
As a result, the creative industries have designed several strategies (moodboards, in-house copyists or work-for-hire contracts) to facilitate the extraction of value from their work. 
Similarly, style transfer systems are designed with the assumption that style is a neutral, discreet object that can be extracted from one source and applied to another. 
Together, these strategies and technologies constitute a socio-technical system designed to optimize value extraction from artistic work, which “metastasizes capitalism” \cite{anarchist}. 

Under the same term ``style'', AI-generated styles now exist alongside human-created ones, a shared social space increasingly resulting in competition between the two. 
This is the ``robust'' quality of style as a boundary object that allows it to be common to several communities. 
However, as we have seen, with regards to artists, AI-generated style is actually a boundary-closing object, it hinders collaboration with them, generating output at odds with their work both in terms of process and results. 
For clients on the other hand, the object becomes boundary-opening, it is aligned with their goals and expectations of what images are for and how to extract value from them.
When left unquestioned as a single, unified object, style becomes the ground on which artists' obsolescence is discussed, another form of human labor that can simply be automated. 
But when analyzed as a multifaceted boundary object, what appeared to be neutral becomes particular, design choices become political, aesthetic results become economic. 
It is thus every researcher's responsibility to be attuned to the boundary they open and close when designing systems.
Our analysis of style transfer contributes to approaches calling for “located accountabilities” of system design \cite{suchman2002located}. Designing systems involves creating and managing boundary objects that facilitate or hinder collaboration between various communities and can enhance power imbalance. 
Researching style cannot be limited to what Donna Haraway \cite{haraway} has called “the god trick of seeing everything from nowhere” but has to be grounded in a critical understanding of the “normative grounds” \cite{power} that generative models create for their users. 

Our study shows that looking at (and working with) style demands a standpoint and that designers should be aware of how, as a boundary object, it “resists any single, totalizing, or universal point of view” \cite{feministhci}. 
Using common sense terms such as ``style'' does not mean every stakeholder is understanding the same thing, which in turn means that the systems we design do not necessarily work for everyone. 
While the model's output failed as a boundary object for creative work for artists, they suggested it might succeed as a boundary object for their clients to extract value from their work, which is two very different things.
Taking the perspectives of commercial artists like illustrators has helped compensate for the cultural blindspot of studying art in isolation from other social systems like economy or politics \cite{child2019working}. 
By analyzing style transfer from the vantage point of the political economy of illustration, we have shown that style, rather than simply about aesthetics, is akin to a form of labor from which value can be extracted. 
We thus relocate discussions of style transfer alongside other works on the exploitation of the ``unwitting labourer'' \cite{morreale2023unwitting} and micro-workers \cite{tubaro2020trainer} required to feed machine learning models. 
In a context where so many Creativity Support Tools might displace workers from their industry, we encourage future HCI research on creative AI to engage directly with questions of labor.
If, as anthropologist Lily Chumley \cite{chumley2016creativity} writes, “style is one way to know you are in capitalism,” style transfer is, under the garb of a creativity support tool, essentially a form of supply-chain optimization.

\section{Limitations and Implications for Future Work}
\label{sec:limitations}
Our study is limited to the perspectives of a small sample of professional illustrators working in editorial, branding and advertising contexts who agreed to have their work fine-tuning a model. We acknowledge that this strong self-selection bias limits the insights and reproducibility of our results.
While this choice was essential to spend enough time with each artist to explain the project, answer concerns, and conduct long interviews and follow-ups, a larger sample would provide more nuanced results. 
We introduced the concept of “boundary objects” to study style, which opens doors for more research on the perceptions of other actors working with style in HCI. 
For the scope of this study, we centered illustrators' perspectives and discussed other actors’ relation to style indirectly through their eyes. 
This was important to foreground illustrators’ narratives in a technology that impacts them most, but our theoretical frameworks would gain from a greater variety of views. 
Finally, we limited our study to one form of style transfer, the fine-tuning of a StableDiffusion model with a LoRa. 
Other systems might provide additional insights into the performance and perception of different style transfers. 
We want to stress that our research did not focus on the performance of a particular model, and the participants’ evaluation of style transfer pertained more to its underlying logic (e.g., Content-Style disentanglement) common to most style transfer tasks.

These limitations are invitations for future research to study style in all its complexity and pluralism. 
While we limited our sample to illustrators in the editorial, advertising, and branding industries, illustrators working in publishing, entertainment, or animation would undoubtedly enrich these perspectives. 

Our study opens several directions for future research. 
We suggest that because art is not separate from other social spheres, especially economy \cite{menger2014economics, becker2023art}, and the creative industries at large rely on individuals commodifying some personal traits (style, voice, appearance, etc.), research in creative AI should always engage with how systems benefit from, align with or disrupt the local power dynamics in a given field. How are other artistic fields (music, acting, etc.) organizing this interface between the personal practice of artists and the value they generate for their industry? How does our study echo more direct forms of AI-mediated extraction of even more personal objects such as actors' bodies or singers' voices? How does a commitment to understanding such dynamics change the way and for whom systems are designed? Conversely, what kinds of socio-technical structure enable artists to use AI tools in their work in their own terms \cite{caramiaux:hal-03762351}?

As we alluded to, style transfer optimizes a view of illustration aligned with the largely noncreative (at least in the romantic sense) managerial class of the creative industry, rather than artists’. Future research at the intersection of creative AI and HCI could therefore focus on this class of workers to study how they shape the supply chain of style as a commodity. How do they interpret style, its source, role, and value? What makes the difference between a system aligned with artists and one aligned with managers? Future research may also want to study artists' uptake of these systems within their own practice, but also in relationship of the political economy of such practice. 
Understanding style as a boundary object opens such directions by localizing it within larger social contexts, and offers HCI researchers a more concrete approach to an otherwise abstract concept.
\section{Conclusion}
The concept of style is incredibly valuable to illustrators, both artistically and economically.
Interestingly, their perception of style in generative AI research often goes unexamined, even when it is central to the systems’ design. 
This oversight has led some designers of generative AI models to assert, hyperbolically, that models can replicate “artistic styles”. While style transfer results might be analogous to style (i.e., they share superficial similarities with it) \cite{imageanalogies}, we’ve shown they are not homologous to it (i.e., they lack its structural substance). 
However, the conflation between the two has created a misleading equivalence between machines' and artists' creations, which has been used to justify illustrators’ displacement from the creative industries\footnote{See for example OpenAI's CTO Mira Murati's infamous statement about generative AI replacing creative jobs that ``shouldn't have been there in the first place.'': \url{https://www.youtube.com/watch?v=Ru76kAEmVfU}}.

We found that style can be seen differently from various perspectives and that for artists, what is generated by style transfer, while related to style, does not amount to it. 
For these reasons, the participants shared that they were not worried about the ability of style transfer to get even close to what they do. 
It is important to stress that their criticisms were not based on the quality of the output but on the underlying logic of style transfer’s design. 

How can we explain, then, that generative AI models are nonetheless disrupting the creative industries? By understanding style transfer’s results as multifaceted boundary objects, we showed that while useless to artists, this shallow copy of style might be enough for their clients. 
For the artists we interviewed, this is nothing new. 
The ability of generative AI models to extract value from their work while not crediting or remunerating them merely follows a long history of similar practices in their industry that predate AI. 
Generative AI, in general, and style transfer, in particular, have often been praised as innovative technological solutions, but for illustrators who try to make a living from their art, they mostly carry on old capitalist problems.

\bibliographystyle{ACM-Reference-Format}
\bibliography{sample-base}

\end{document}